\documentclass{ws-ijmpa}

\usepackage{xurl}
\usepackage[super]{cite}
\usepackage{xcolor}
\usepackage[verbose,hypertexnames=false]{hyperref}
\hypersetup{colorlinks=false,allbordercolors=blue,pdfborderstyle={/S/U/W 1}}

\begin{document}

\markboth{T.\ Mengoni}{GW generation in unified dark sector warm inflation}

%
\catchline{}{}{}{}{}
%

\title{Continuous Bogoliubov formalism for gravitational-wave generation in a unified dark sector warm inflation }

\author{Tommaso Mengoni
}

\address{University of Camerino, Via Madonna delle Carceri, Camerino, 62032, Italy\\
INAF - Osservatorio Astronomico di Brera, Milano, Italy\\
Istituto Nazionale di Fisica Nucleare (INFN), Sezione di Perugia, Perugia, 06123, Italy\\
tommaso.mengoni@unicam.it}

\maketitle


\begin{abstract}

In this work, we review gravitational-wave generation in a two-scalar-field cosmological model. The framework relies on a two-field scenario in which warm inflation is unified with the dark sector within a single theoretical description. It has recently been shown that this setup leads to a gravitational-wave spectrum potentially detectable by future experiments. Here we also demonstrate that the resulting spectrum is largely independent of the $\phi$ field initial condition, thereby extending previous findings.

\end{abstract}

detection 
\ccode{PACS numbers: 04.30.-w, 95.36.+x, 95.35.+d, 98.80.Cq, 03.50.-z}


\section{Introduction}\label{Section 1}

A new era in cosmology began with the first direct detection of gravitational waves \cite{Abbott:2016}. 
So far, gravitational-wave astronomy has primarily observed binary stellar-mass black hole systems. 
In the near future, next-generation detectors are expected to probe a wider range of phenomena \cite{Bailes:2021}.

Among the most compelling targets is the generation of primordial gravitational waves in the early stages of the Universe, which could provide valuable information about inflation and the transition to the radiation-dominated era \cite{Roshan:2025}.

Because of the lack of direct observational constraints on these epochs, several theoretical models have been proposed.
In this context, unification scenarios are particularly interesting\footnote{For broader discussions on unification models, see, e.g., Refs. 
\citen{Aviles:2011sfa, Dunsby:2016lkw, Capozziello:2018mds, Boshkayev:2019qcx}.}, as they aim to describe the dark sector and inflation within a single framework \cite{Henriques:2009, Luongo:2018lgy, Luongo:2024opv, Paliathanasis:2025mvy}. 
A triple unification scheme can be realized in a two-scalar-field cosmological model that also considers a warm inflation scenario \cite{sa-2020a, Sa:2020, Sa:2021}.
In Refs. \citen{Luongo:2026 ,Mengoni}, the generation of gravitational waves in this model was extensively studied. 
In this contribution, we review those results, which predict spectra potentially detectable by next-generation detectors, and we further investigate the role of the initial conditions of the dark energy field.

This proceeding is organized as follows. 
In Sect.~\ref{Section 2}, we briefly review the two-scalar-field cosmological model and the continuous Bogoliubov coefficient formalism used to compute the gravitational-wave spectra. 
In Sect.~\ref{Section 3}, we present and discuss the resulting spectra. 
Finally, in Sect.~\ref{Section 4}, we summarize our findings and outline future perspectives.


\section{The cosmological model and the continuous Bogoliubov formalism}\label{Section 2}

The two-scalar-field cosmological model is characterized by the fields $\xi$, playing the role of the inflaton and subsequently of dark matter, and $\phi$, representing dark energy \cite{Sa:2020, Sa:2021, Luongo:2026, Mengoni}. The corresponding action is given by
\begin{equation}
     S = {}  \int d^4x \sqrt{-g} \bigg[
  \frac{R}{2\kappa^2} - \frac12 (\nabla \phi)^2 - \frac12 e^{-\alpha\kappa\phi} (\nabla \xi)^2
  - e^{-\beta\kappa\phi} V(\xi) \bigg]\,,
 \label{action-2SF}
\end{equation}
where $\kappa=\sqrt{8\pi}/m_p$ ($m_p$ is the Planck mass) and $\alpha,\,\beta$ are free parameters of the model that determine the non-minimal coupling between the scalar fields. 
In particular, they have been fixed through statistical analyses\cite{Luongo:2022} as
\begin{equation}
    \alpha = 0.36^{+0.18}_{-0.26}\,,\quad{\rm and}\quad
\beta= 0.01^{+0.34}_{-0.24}\,,
\label{v-alphabeta}
\end{equation}
while the inflationary potential can be set as $V(\xi)=V_a+\frac12 m^2 \xi^2$.
The warm inflationary scenario\cite{Berera:1995, kamali:2023} is modeled in a phenomenological way, parameterizing the dissipation coefficients $\Gamma_{\xi,\phi}$ via the parameters $q$, $p$, and $f$
\begin{align}
 \Gamma_{\xi,\phi} = f_{\xi,\phi} \times \left\{
    \begin{aligned}
     & T^p,
     & T\geq T_\texttt{E},
    \\
     & T^p  \exp \left[ 1- \left( \frac{T_\texttt{E}}{T} \right)^q \right],
     & T\leq T_\texttt{E},
    \end{aligned}
 \right.
 \label{gammas}
\end{align}
with $T_E$ the temperature of the radiation bath at the end of inflation.
To account for gravitational-wave generation, the tensor perturbations $h_{ij}$ to the Friedmann-Lema\^{i}tre-Robertson-Walker metric are considered 
\begin{equation}
    ds^2=a^2(\eta) \{ -d\eta^2 + [ \delta_{ij} + h_{ij}(\eta,\textbf{x}) ] dx^i dx^j \},
\end{equation}
where $\eta$ is the conformal time.
The continuous Bogoliubov coefficients formalism for gravitational waves consists of interpreting tensor perturbations as gravitons\cite{Mendes:95, Sa:2008, Sa:2012, Luongo:2026, Mengoni}, $\chi(\eta,\textbf{k})$, such that they can be quantized as
\begin{equation}
h_{ij}(\eta,\textbf{x})=\kappa\sum_{p=1}^2 \int \frac{d^3k}{(2\pi)^{3/2}a(\eta) \sqrt{2k}}\left[a_p(\eta,\textbf{k}) \varepsilon_{ij}(\textbf{k},p) e^{i\textbf{k}\cdot\textbf{x}}\chi(\eta,\textbf{k}) + \mbox{h. c.}\right]\,,
\end{equation}
where $\textbf{x}$ is the spatial-coordinates three-vector, $\textbf{k}$ is the comoving wave-number three-vector, $k=|\textbf{k}|=a\, \omega$, $\omega$ is the angular frequency, $p$ runs over the two polarizations of the gravitational waves, $a_p$ is the annihilation operator, $\varepsilon_{ij}$ is the polarization tensor. Within this framework, their evolution is encoded in the evolution of the ladder operator that follows the Bogoliubov transformation
\begin{equation}
a_p(\eta,\textbf{k})=\alpha_k(\eta,k)A_p(\textbf{k})+\beta_k^*(\eta,k)A_p^\dagger(\textbf{k}),
\end{equation}
where the Bogoliubov coefficients can be written in terms of the continuous functions $X$ and $Y$, $\alpha_k=\frac12(X+Y)\exp\{ik(\eta-\eta_i)\}$ and $\beta_k=\frac12(X-Y)\exp\{-ik(\eta-\eta_i)\}$, such that 
\begin{subequations} \label{SODE-XY}
\begin{align}
    X^\prime&=-ikY, \label{Xprime}
\\
    Y^\prime&=-\frac{i}{k} \left( k^2-\frac{a^{\prime\prime}}{a} \right) X, \label{Yprime}
\end{align}
\end{subequations}
with initial conditions $X_i=Y_i=1$, \emph{i.e.} no gravitons produced.
From the analogy to gravitational particle production \cite{Parker:69, Belfiglio:2024swy}, the number of gravitons created is given by $\langle N_k(\eta) \rangle = |\beta_k|^2$, and, consequently, the gravitational-wave spectral energy density parameter reads
\begin{equation}\label{eq spectrum}
\Omega_{\texttt{GW}}(\omega_0)=\frac{8\hbar G}{3\pi c^5 H^2_0} \omega_0^4 \,(|\beta_{k}|^2)_0\,,
\end{equation}
where the label $0$ denotes the quantities corresponding to the current time\footnote{Particle production in cosmological scenarios is a widely studied phenomenon, in several applications and variants such as in Refs. \citen{Belfiglio:2022qai, Belfiglio:2023rxb}.}.


\section{The gravitational-wave spectra}\label{Section 3}

\begin{figure}[ht]
    \centering
    \includegraphics[scale=1.]{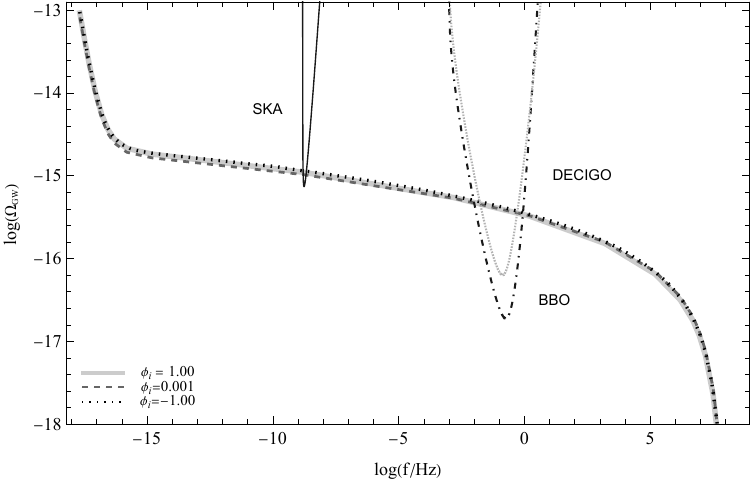}
    \caption{Full gravitational-wave energy spectrum for the addressed cases, superimposed on the sensitivity curves of planned next-generation ground- and space-based gravitational-wave detectors, BBO, SKA, and DECIGO \cite{Schmitz:2020syl, repository}.}
    \label{fig:1}
\end{figure}

We then compute the gravitational-wave spectrum for all possible frequencies\footnote{The spectra are presented in terms of the frequency $f$, related to the angular frequency $\omega_0$ by the relation $\omega_0=2\pi f$.} spanning from the one corresponding to the horizon at the end of inflation ($f\simeq10^{-18}{\rm\,Hz}$) towards the one corresponding to the current horizon ($f\simeq10^{8}{\rm\,Hz}$). 
We adopt the same base scenario as in Ref. \citen{Luongo:2026}, where the parameters of the model correspond to $\alpha = 0.36$ and $\beta= 0.01$, with $H_0= 1.17 \times 10^{-61} \, m_\texttt{P}$, as well as $p=1$, $q=2$, $f_\xi=f_\phi=2$, $V_a=1.1\times10^{-123} \,m_\texttt{P}^4$ and $m = 10^{-5} \,m_\texttt{P}$, in addition to the initial conditions $\xi_i=0.75 \, m_\texttt{P}$, $\phi_i=10^{-3} \,m_\texttt{P}$, $\xi_{u,i}=10^{-2} \,m_\texttt{P}$, $\phi_{u,i}=10^{-5} \,m_\texttt{P}$, and $\rho_{\texttt{R},i}=0.25\times10^{-12}\, m_\texttt{P}^4$\footnote{The label $u$ denotes the derivative with respect to the variable $u=-\ln(a_0/a)$, which is chosen because of computational advantages.}.
These values are chosen in order to satisfy $70$ e-folds of inflation and in accordance with observational constraints\cite{planc}.  
Specifically, the initial condition on the field $\phi$ is chosen such that dark energy is initially subdominant compared to the other components. This constraint allows for multiple viable choices.
In this regard, we also considered the cases where $\phi_i=\pm1.00$, and, in Fig. \ref{fig:1}, the spectra are shown.
The resulting energy density parameters $\Omega_{GW}$ yield comparable results, proving that the gravitational-wave production is largely independent of that specific initial condition.
Moreover, Fig. \ref{fig:1} confirms that such a spectrum could be probed by future detectors such as Big Bang Observer (BBO), Square Kilometre Array (SKA), and Deci-Hertz Interferometer Gravitational Wave Observatory (DECIGO).

\section{Conslusions and perspectives}\label{Section 4}

In this contribution, we started by briefly reviewing the recent works in Refs. \citen{Luongo:2026, Mengoni}.
Specifically, we explored a triple-unification cosmological model consisting of a two-scalar-field scenario within a warm inflation epoch.  
This framework is used to analyze the generation of primordial gravitational waves by means of the continuous Bogoliubov coefficients method.
Then, we extended the results previously obtained by showing that the resulting spectrum is largely independent of the initial condition of the dark energy field $\phi$.
Moreover, the spectra appear to be within the sensitivity range of future detectors, thereby reinforcing the role of gravitational-wave astronomy as a crucial tool for investigating the very early Universe.
Within this perspective, further studies on these phenomena are needed.


\section*{Acknowledgments}

TM gratefully acknowledges Paulo M. S\'a for the collaboration on Refs. \citen{Luongo:2026, Mengoni}, on which this contribution is largely based.

\section*{ORCID}

\noindent Tommaso Mengoni - \url{https://orcid.org/0009-0007-2955-3521}

\end{document}